\documentstyle[pre,twocolumn,aps,psfig]{revtex}
\begin{document}
\preprint{}
\draft
\twocolumn[\hsize\textwidth\columnwidth\hsize\csname@twocolumnfalse%
\endcsname
\title{Discrimination between Healthy and Sick Cardiac Autonomic
Nervous System by Detrended Heart Rate Variability Analysis}
\author{Y.C Ashkenazy$^{a,b}$, M. Lewkowicz$^{c,a}$, J. Levitan$^{c,a,d}$,
S. Havlin$^{a,b}$,\\ K. Saermark$^d$, H. Moelgaard$^e$ and 
P.E. Bloch Thomsen$^f$\\ (a) Dept. of Physics, Bar-Ilan University, Ramat-Gan, 
Israel\\ (b) Gonda Goldschmied Center, Bar-Ilan University, Ramat-Gan, Israel\\
(c) Dept. of Physics, College of Judea and Samaria, Ariel, Israel\\ 
(d) Dept. of Physics, The Technical University of Denmark,\\ Lyngby, Denmark\\ 
(e) Dept. of Cardiology, Skejby Sygehus, Aarhus University Hospital,\\
Aarhus, Denmark\\ 
(f) Dept of Cardiology, Amtssygehuset i Gentofte,\\
Copenhagen University Hospital, Denmark \\}
\date{\today}
\maketitle
\begin{abstract}
{
Multiresolution Wavelet Transform and Detrended Fluctuation Analysis
have been recently proven as excellent methods in the analysis of
Heart Rate Variability, and in distinguishing
between healthy subjects and patients with various dysfunctions of the
cardiac nervous system. We argue that it is possible to obtain
a distinction between healthy subjects/patients of at least similar
quality by, first, detrending the time-series of RR-intervals by subtracting
a running average based on a local window with a length of around 32 data
points, and then, calculating the standard deviation of the detrended 
time-series. The results presented here indicate that the analysis can be
based on very short time-series of RR-data (7-8 minutes), which is a
considerable improvement relative to 24-hours Holter recordings.
}
\end{abstract}
\vspace{0.5cm}
]
\narrowtext
\newpage
\section{Introduction}
Measurements of Heart Rate (HR) and evaluation of its rythmicity have been
used for a long time as a simple clinical indicator. Research from the last
decade has indicated that a quantification of the discrete beat to beat
variability in HR - the heart rate variability (HRV) - might be a possible
prognostic indicator of risk associated with a large variety of diseases, 
behavioral disorders, mortality and also ageing. For example, independent
of other established risk factors, depressed HRV has been shown to be a
powerful predictor of cardiac events after myocardial infarct. It is therefore
of great importance to establish a measure of HRV and to classify the HRV of
different pathological cases in order to discriminate the healthy HRV profile
from that for patients at risk \cite{{Moelgaard95},{Malik98},{Moelgaard94},{Wolf78}}. 
It is an open question in the
literature if one needs long time series (24 hour ECG Holter data series 
\cite{Klingenheben98})
or whether short time series (ca. 5 minutes \cite{Faber96}) do suffice in 
producing a
reasonable clear separation between healthy and sick individuals. This question
is probably tied up with the quality of the ECG recording, i.e. the signal
to noise ratio.

In physiological systems one can recognize different behaviours at different
time scales. For example, consecutive heart beats will occur more or less with 
the same beat-to-beat interval (the mean HR), which can be defined as a small 
time scale. Other time scales can be defined by the sleep/wake periods. On
these larger time scales one can identify a different heart rate and a
different heart rate variability during the hours of sleep and  during the 
hours of awakeness. The DFA (Detrended Fluctuation Analysis) \cite{Peng95}
 and the
DWT (Discrete Wavelet Transform) 
\cite{{Ivanov96},{Thurner98},{Ashkenazy98},{Daubechies92},{Strang96},{NR},{Aldoubri96},{Akay97}} have been shown to be 
succesful
ways of analyzing the HRV. Basically, these methods explore the low and the
high frequency behaviour of the signal at different time-scales by applying
windows of varying lengths. Thus the DWT was used \cite{Thurner98} 
to analyze data from RR
measurements and calculating the standard deviation of the transformed data.
This standard deviation of the Wavelet coefficients serves as a 
characterization of the HRV during the period of measurement and it was 
shown that the
method discriminates between healthy and sick individuals. Thuerr {\it et al}
\cite{Thurner98} observed a complete separation of the two groups for window 
sizes $2^4$
and $2^5$, where the exponent indicates the window scale. Further, this method
was used by our group \cite{Ashkenazy98} on a different set of data, and the 
separation
mentioned was found not to be complete (see also \cite{Amaral98}). In order 
to improve the method a
filtering algorithm was constructed and the standard deviation of the filtered
time-series now resulted in a complete separation between healthy and sick
subjects. We emphasize that the diagnostic virtue of the DFA and DWT methods
apparently is due more to the right choice of window size than to the actual
method of transformation. Both of the two methods point to a typical time
scale of $2^4$ to $2^5$ equivalent to a window size of 16 to 32 heart beats.
Thus in the DFA method \cite{Peng95} one observes a crossover point  for a 
window size
around n=16 heart beats where an abrupt change in the slope of their 
\cite{Peng95}
$F(n)$ vs. $n$ curve occurs. 

In this study we wish to see if the existence of this
time scale can be utilized for the analysis of HRV of short term ECG 
recordings.

\section{The Detrended Time Series}
\begin{figure}[thb]
\psfig{figure=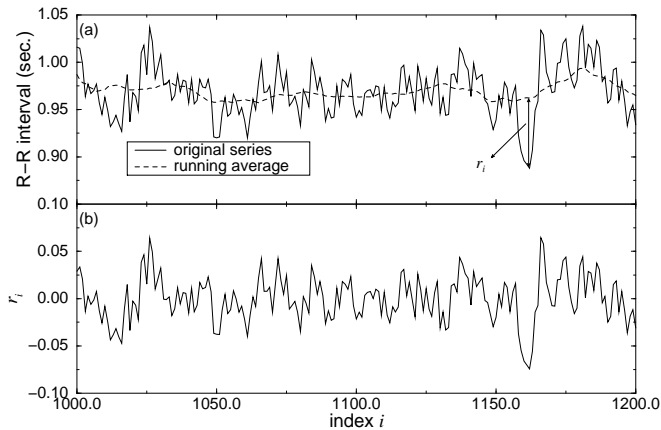,height=5.7cm,width=10cm,angle=-90}
\caption[]{\label{fig1}
(a) Segment of RR-interval data versus beat number (solid curve)
and running average based on a local window of 32 heart beats (broken 
curve). (b) Detrended curve, i.e. the difference between
solid curve and broken curve in top panel.
}
\end{figure}

As mentioned in the Introduction both the DFA analysis and the DWT analysis
suggest an intrinsic window of scale $m=4$-5, i.e. a window consisting of 16-32
heart beats. In this section we utilize this to perform a detrending in the
following way. First, from the time-series of the raw RR-data 
\footnote{An RR interval is the time difference between two consecutive 
pronounced peaks - the R peaks - of the ECG recording.} a running
average is constructed using an interval-length of $2^m$
\footnote{In this study we confined ourselves to an interval length of
$2^m$, although any interval length can be chosen.}. Next, the running 
average is subtracted from the original RR-data time series. For $m=5$ this 
procedure
is illustrated in Fig. \ref{fig1}a, where the solid curve represents the raw 
RR-data and
the dashed curve represents the running average.The difference between the two
curves is denoted by $r_i$ and is shown in Fig. \ref{fig1}b. This resultant 
time-series
$r_i$ is here called the detrended time-series (DTS) and represents the 
fluctuations
with respect to the local average. It is hoped that this procedure at least
partly will remove noise and slow oscillations which should not directly affect
short term HRV \cite{{Moelgaard95},{Moelgaard94}}.

\begin{figure}
\psfig{figure=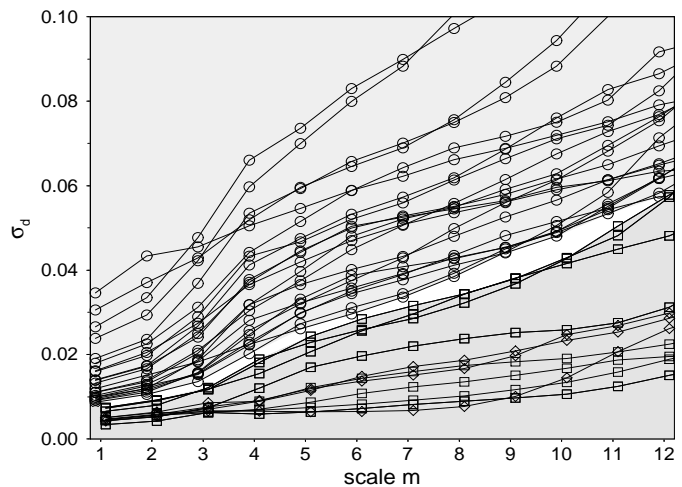,height=6.5cm,width=10cm,angle=-90}
\caption[tbp]{\label{fig2}
Standard deviation (in seconds) of the detrended series for a group of
33 subjects versus the scale factor of the local window used in the
detrending. Healthy subjects: Circles, Diabetics: Squares and Heart
patients: Rhombohedra. The three topmost diabetics have been shown
with a special marking.
}
\end{figure}

\begin{figure}
\psfig{figure=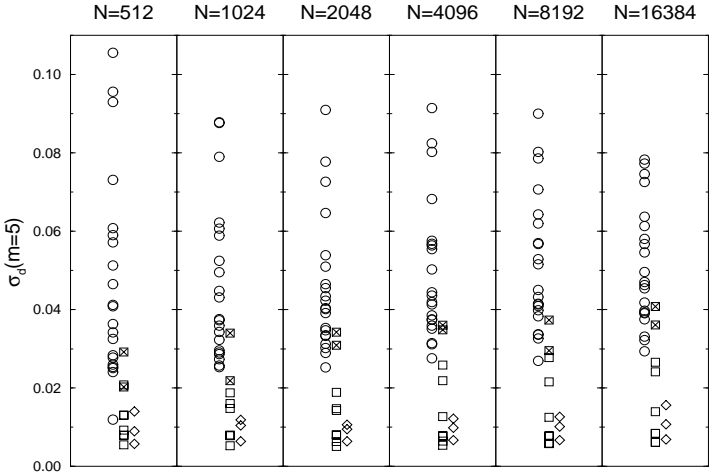,height=7cm,width=10cm,angle=-90}
\caption[tbp]{\label{fig3}
The standard deviation (in seconds) of the detrended RR-data time-series
versus the length of the time series analyzed. Symbols: Same as in Fig. 
\ref{fig2}
}
\end{figure}

The standard deviation $\sigma_d$ of the detrended time-series (DTS), using
a detrending window of scale m, includes now only the behaviour of relevant
small time scales and may thus be considered a measure of the HRV. To evaluate
the discriminating capabilities of $\sigma_d$ we examined RR-data for a group
of 33 subjects (the same data group as in ref. \cite{Ashkenazy98}) consisting 
of 21 healthy
subjects, 9 diabetics and 3 heart patients including one heart transplanted
patient. Thus we calculate $\sigma_d$ for a time-series consisting of
$2^{16}=65536$ data points, corresponding to approximately 16 hours of measured
ECG data, and for the scale values $m=1$-12 for the detrending window. The
smallest length of the detrending window is thus 2 and the largest 4096. The
results are shown in Fig. \ref{fig2}, which is the analogous of Figs. 
\ref{fig2} and \ref{fig3} in
ref. \cite{Ashkenazy98}; the latter two figures. show, respectively, the 
standard deviation of
the wavelet coefficients and of the filtered time series. In the present 
Fig. \ref{fig2}
one notes a clear separation between the group of healthy subjects (circles)
on the one hand, and the groups of diabetics (squares) and heart patients 
(rhombohedra) on the other hand. However, one also notes from this figure 
 that 3 of
the diabetics (the three topmost) with as much justification
could have been included in the group of healthy subjects thus displacing the
separation region for $\sigma_d$ towards lower values. For systematic reasons
we have chosen the separation region shown in Fig. \ref{fig2}.

In Fig. \ref{fig2} the largest separation betwen the healthy subjects and the 
two other
groups is found for the scale $m=8$-11, whereas for the DWT analysis the
largest separation was found for the scale $m=4$-6, see ref. \cite{Ashkenazy98} 
(see also ref. \cite{Amaral98} for discussion of the dependence of the method)
and for the DFA
analysis the crossover point for the fractal slope was found for the scale
$m=4$, see ref. \cite{{Peng95},{Amaral98}}. It should be noted, however, that the 
crossover point in the
DFA analysis is not a sharply defined point, rather the change in fractal slope
takes place in a gradual way. On this basis we conclude that the three methods
DTS, DWT and DFA yield equivalent estimates for the scale of a characteristic
window and in the following we use the scale $m=5$
\footnote{The maximum separation depends on the size of the interval;
we have found that $m=5$ offers the optimal scale for small interval lengths.}. 

\section{Estimation of Interval Length}
In the preceeding section we used an RR-data time-series corresponding to 16
hours of ECG measurements. Clinically it is of course of importance to be able
to use as short time-series as possible. In this section we use the DTS method
to examine whether or not short time series can be used in order to distinguish
between the group of healthy subjects and the groups of diabetics and heart
patients. Specifically we choose time-series of lengths 512, 1024, 2048, 4096,
 8192 and 16384 data points (RR-intervals). To make sure that all data are
 collected under
similar conditions we have only used data points from the sleep period starting
at the initial time 1 a.m. For the various lengths of the time-series we used
the same window scale $m=5$ and the resulting $\sigma_d$ is shown in 
Fig. \ref{fig3}. One
notes from the figure that for all time-series lengths - including the very 
short one of 512 measurements corresponding to 7-8 minutes measuring time - an
almost complete separation between the different groups (healthy, diabetic and
heart patients) is obtained. The only exception is that 2-3 diabetics (marked
squares) overlap the group of healthy subjects. As noted in the preceeding 
section, these diabetics appear to fall in a group for themselves and can 
with some
justification be regarded as belonging to the healthy group of subjects, i.e.
as being of no immediate heart risk. Moreover, the entire group of heart 
patients falls into the lower range of the $\sigma_d$ scale, 
$\sigma_d \le 0.015$
where only a few of the diabetics are found. We remark, that the small
number of diabetics and heart patients allows for no definitive conclusion, but
we do argue that there are strong indications that even a small length of the
RR-data time-series, say 512 measurements, do allow for an almost complete
separation between healthy subjects and heart patients/diabetics.
       
\begin{figure}
\psfig{figure=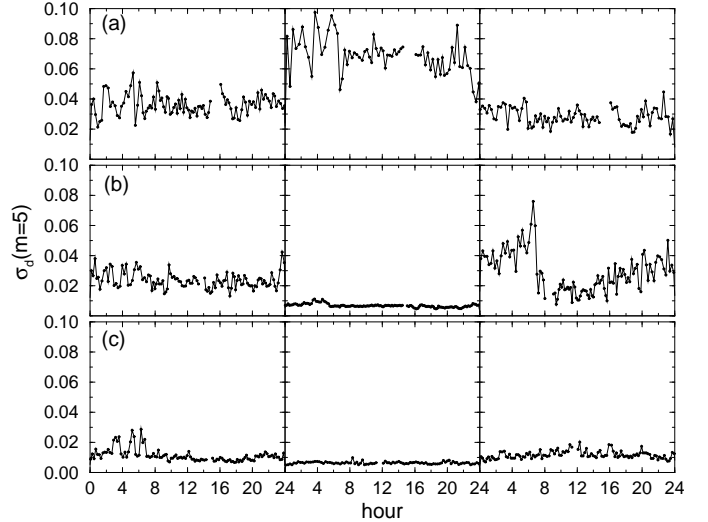,height=7cm,width=10cm,angle=-90}
\caption[tbp]{\label{fig4}
The daily variation of the standard deviation, calculated for data
segments of length 512, versus the time location for the segment. Three
representative recordings for each of the groups: (a) Healthy subjects,
(b) Diabetics and (c) Heart patients. Again the standard deviation was
calculated for the detrended time-series and the scale $m=5$, i.e. 32
heart beats, was used in the detrending procedure.
}
\end{figure}
\section{Daily Variation of HRV}
\begin{figure}
\psfig{figure=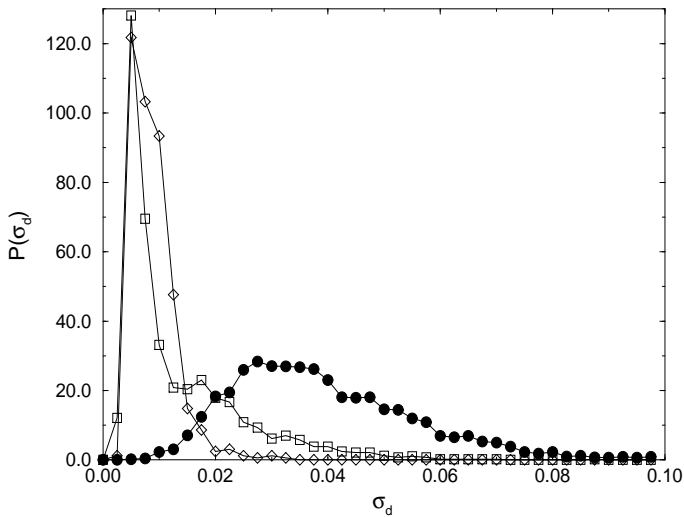,height=7cm,width=10cm,angle=-90}
\caption[tbp]{\label{fig5}
Histograms for the daily variation of the standard deviation (segment
length and scale m as in Fig. \ref{fig4}) shown for the three groups: Healthy
subjects (circles), Diabetics (squares) and Heart patients (rhombohedra).
}
\end{figure}

\begin{figure}
\psfig{figure=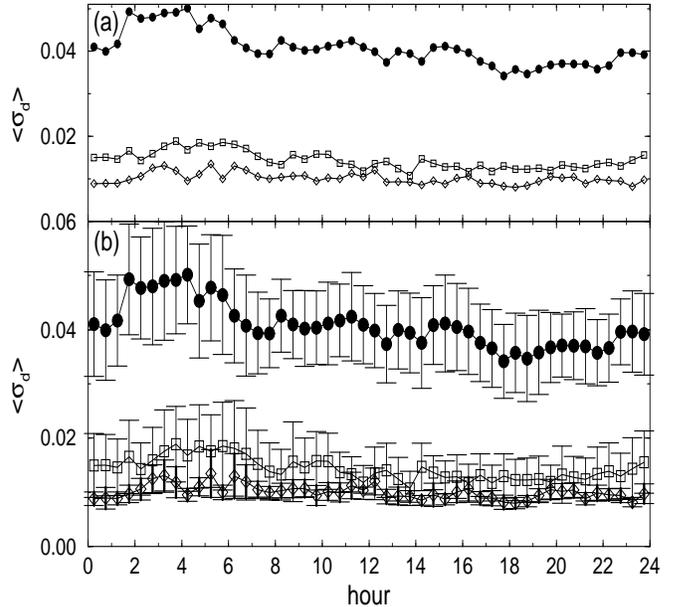,height=9cm,width=10cm,angle=-90}
\caption[tbp]{\label{fig6}
(a) The group average of the standard deviation (parameters as
in Fig. \ref{fig4}.) versus the time location for the data segment. Healthy
subjects: circles, Diabetics: squares and Heart patients: rhombohedra.
(b) Same as (a), but with the standard deviation across
the group indicated by vertical bars.
}
\end{figure}

We next study the daily variation of the standard deviation of the detrended
24 hour ECG time series. The full data series was divided into smaller segments
and in agreement with the conclusion of the preceeding section we choose a
segment length of 512 measurements, not including artefacts. 
Althoguh the daily variation was calculated for all 33 subjects
(the same group as in ref. \cite{Ashkenazy98}) we here display only 9 
representative results.
Thus in Fig. \ref{fig4} we show 3 healthy subjects (top panel, (a)), 
3 diabetics (middle panel, (b)) and 3 heart patients (lower panel, (c)) 
and in all cases $\sigma_d$ is
plotted versus the time position of the data segment. From Fig. \ref{fig4}
one notes, first, that the average value (across 24 hours) of $\sigma_d$ is far
smaller for the heart patients than for the healthy subjects, whereas the
diabetics form a more varied population: one (central recording) clearly 
similar
to recordings for heart patients, one (middle panel, third recording) 
reminiscent of recordings for healthy patients and one (middle panel, first
recording) appears to be an in-between case. Secondly, the fluctuation of
$\sigma_d$ during the 24 hours of observation appears to follow the same
pattern as just described.
We illustrate the above features in two other ways. In Fig. \ref{fig5} 
we show
the histogram for the daily (24 hours) variation of $\sigma_d$ for the 3
groups: 21 healthy subjects (circles), 9 diabetics (squares) and 3 heart 
patients (rhombohedra). For the healthy subjects the maximum of the histogram,
i.e. the most probable value of $\sigma_d$, is well separated from the
maximum of the two other histograms and there is very litlle overlap with the 
histogram for the heart patients but some overlap with the histogram for
the diabetics group. The histograms thus clearly distinguish between healthy
and sick subjects, but the statistics is too poor to distinguish between the
two patient groups. Finally, in Fig. \ref{fig6} we show the group average of 
$\sigma_d$
for each of the three groups versus the time position for the relevant data
segment. In Fig. \ref{fig6}b we have shown the standard
deviation (across the group) of the group average of $\sigma_d$. From
Fig. \ref{fig6}a it follows that the healthy group is separated from 
the heart
patient group by a factor of around 4 in the average of $\sigma_d$, a 
conclusion supported by the standard deviations shown 
in Fig. \ref{fig6}b. It is
also tempting to draw the conclusion that the diabetics group is separated from
the heart patient group by 25-50\%, however the standard deviations shown
in the bottom panel do not allow for a definite conclusion. It also appears 
from Fig. \ref{fig6}, that the group average value of $\sigma_d$ is larger 
during the
sleep period, say from 2 to 7 hours, for the healthy group and partly also for 
the diabetics group (see also \cite{Ivanov98}). For the healthy group the 
difference between the sleep and wake period is around 0.01 s. 

\section{Conclusion}
In this paper we have focused on the standard deviation $\sigma_d$ of 
detrended RR-data time-series using a detrending window with a length $2^5$
measurements of RR-data, this value being indicated by results from
DWT and DFA analyses, see refs. \cite{{Ashkenazy98},{Peng95}}. Our results suggest that even
a short time-series of 512 data points, i.e. 7-8 minutes of measurements,
suffices to distinguish between healthy subjects and patients (heart disease,
diabetics) . The same kind of analysis can of course be performed on the raw, 
not detrended data \cite{Wolf78}. We have done so, but find that using 
detrended data series
 are more succesful. We note, that even if the length of the RR-data 
time-series would
have to be increased to, say, 2048 measurements corresponding to app. half an
hour of ECG measurements, this would still from a clinical point of view
represent a substantial advantage relative to 24 hours of ECG measurements. 

The authors are grateful to the Danish-Israel Study Fund in memory of Josef 
and Regine Nachemsohn. Y.A. acknowledges support from The Sharbat Foundation.

\end{document}